\documentclass[english,aps,twocolumn,showpacs,amsmath,amssymb,prb,superscriptaddress]{revtex4-1}
\usepackage[T1]{fontenc}
\usepackage[latin9]{inputenc}
\usepackage{color}
\usepackage{amstext}
\usepackage{amssymb}
\usepackage{graphicx}
\usepackage{epstopdf}

\makeatletter

\@ifundefined{textcolor}{}
{%
 \definecolor{BLACK}{gray}{0}
 \definecolor{WHITE}{gray}{1}
 \definecolor{RED}{rgb}{1,0,0}
 \definecolor{GREEN}{rgb}{0,1,0}
 \definecolor{BLUE}{rgb}{0,0,1}
 \definecolor{CYAN}{cmyk}{1,0,0,0}
 \definecolor{MAGENTA}{cmyk}{0,1,0,0}
 \definecolor{YELLOW}{cmyk}{0,0,1,0}
}

\usepackage[caption=false]{subfig}

\@ifundefined{showcaptionsetup}{}{%
 \PassOptionsToPackage{caption=false}{subfig}}
\usepackage{subfig}
\makeatother

\usepackage{babel}
\begin{document}

\title{Time evolution of localized states in Lieb lattices}

\author{J. D. \surname{Gouveia}}
\thanks{These two authors contributed equally to the work.}

\author{I. A. \surname{Maceira}}
\thanks{These two authors contributed equally to the work.}

\author{R. G. \surname{Dias}}

\affiliation{Department of Physics, I3N, \\
 University of Aveiro, \\
 Campus de Santiago, Portugal}

\date{\today}
\begin{abstract}
We study the slow time evolution of localized states of the open-boundary Lieb lattice when a magnetic flux is applied perpendicularly to the lattice and increased linearly in time. In this system, Dirac cones periodically disappear, reappear and touch the flat band as the flux increases. We show that the slow time evolution of a localized state in this system is analogous to that of a zero-energy state in a three-level system whose energy levels intersect periodically and that this evolution can be mapped into a classical precession motion with a precession axis that rotates as times evolves. Beginning with a localized state of the Lieb lattice, as the magnetic flux is increased linearly and slowly, the evolving state precesses around a state with a small itinerant component and the amplitude of its localized component oscillates around a constant value (below but close to 1), except at multiples of the flux quantum where it may vary sharply. This behavior reflects the existence of an electric field (generated by the time-dependent magnetic field) which breaks the C$_4$ symmetry of the constant flux Hamiltonian.
\end{abstract}


\maketitle

\section{Introduction}

In flat-band systems, there is a high energy degeneracy associated with the existence of localized states (i.e. electrons trapped in a small region of a lattice due to destructive wave function interference). Recent interest in this area has arisen\cite{Goldman2011,Leykam2016} due to experimental realizations of flat-band systems using arrays of optical waveguides\cite{Mukherjee2015,Vicencio2015}, exciton-polarization condensates\cite{Masumoto2012,Baboux2016}, and cold atoms in optical lattices\cite{Taie2015}. Known lattices with flat bands include the Lieb\cite{Lieb1989}, Mielke\cite{Mielke1992} and Tasaki\cite{Tasaki1992} lattices and there are methods of generating lattices of nearly arbitrary geometry which have these localized states when the hopping constants obey certain relations\cite{Flach2014,Dias2015}.

These systems can be separated in two classes in what concerns the behavior of their flat band in the presence of an external magnetic field. In particular, the Mielke and Tasaki lattices do not display flat bands for finite magnetic flux. In contrast, lattices of the Lieb's class are flat-band robust in that they retain a flat band when a magnetic field is applied perpendicularly to the lattice. However, the introduction of magnetic flux requires that localized states occupy at least two plaquettes\cite{Lopes2011} and therefore, the flat band subspace as a whole evolves in the Hilbert space as the magnetic field is increased. In the case of the Lieb tight-binding~(TB) model, the band structure has a Dirac point at $\mathbf{k}=(\pi,\pi)$. This model under an evolving magnetic field creates an interesting theoretical scenario: i) a flat band and Dirac bands that touch at the Dirac point when the magnetic flux per plaquette is a multiple of the flux quantum; ii) a perturbation such that the flat band persists, and the Dirac cones disappear and reappear periodically as the perturbation varies.

In this paper, we study the slow time evolution of localized states in the scenario described above. As stated by the adiabatic theorem \cite{Born1928}, if the evolution of the perturbation (magnetic field) is slow enough, the evolving state, initially an eigenstate, is expected to closely remain an instantaneous eigenstate of the Hamiltonian at any time, as long as there is an energy difference between that eigenstate and the rest of eigenstates. Since energy levels periodically cross the flat band, this time evolution will periodically leave the adiabatic regime close to the crossing instants.

One of the questions we wish to answer is: near the energy crossing instants, can we picture the flat band system as a three-level system with one zero-energy (flat-band) state and two finite-energy ones? The motivation for this question is two-fold. First, if one considers a finite-size tight-binding lattice, the Dirac cones are replaced by discrete levels and, as the perturbation is increased, two of these levels cross the flat-band level. Second, the application of the perturbation to the flat-band tight-binding system introduces Hamiltonian terms that couple each dispersive state with flat-band states. However, as the flat band is degenerate, one can rotate the basis of the flat-band subspace in a way such that the perturbation couples the dispersive state with only {\it one} localized state (in analogy with what was done in Ref.~\onlinecite{Lopes2014}). We find that if the magnetic flux is applied linearly and slowly, the localized component of the evolving state oscillates around a constant value, except at energy crossing instants where it varies sharply. This reflects an intricate precession behavior of the evolving state around a state with a small itinerant component. Such a behavior is also found in the case of a three-level system whose energy levels intersect periodically. This precession behavior reflects the existence of an electric field in the Lieb lattice (generated by the time-dependent magnetic field) which breaks the C$_4$ symmetry of the constant flux Hamiltonian.

The paper is organized in the following way. First, we consider the tight-binding Hamiltonian of the Lieb lattice in the presence of magnetic flux (also called $t$-$\phi$ model). When the magnetic flux per plaquette is a multiple of the flux quantum, the flat band has two extra states, which we label $\vert \varepsilon_\pm \rangle$ states. We then analyze the time evolution of one of its localized eigenstates, starting at a certain initial magnetic flux and then varying the magnetic flux linearly and slowly. 
Secondly, we study a toy three-level system with a time-dependent Hamiltonian consisting of one zero-energy eigenstate and two finite-energy ones, whose energy periodically crosses the zero-energy line. We study the slow time evolution of the zero-energy state and find that it is equivalent to a classical precession motion, but with a nutation-like oscillation of the zero-energy component due to the rotation of the Hamiltonian eigenbasis. This precession motion may lead to huge long-time modifications of the state if, when the level crossings occur, the precession axis rotates with finite angular velocity.
Thirdly, we show that analogous precession behavior is found in the evolution of any localized state of the Lieb lattice under time-dependent magnetic flux. In this case, since it is this time-dependence that leads to the rotation of the precession axis, we can also say it is a consequence of the electric field generated by the evolving vector potential.
\section{Lieb lattice under magnetic flux}
\label{sec:lieb}
\begin{figure*}[th]
\centering
\includegraphics[width=1 \textwidth]{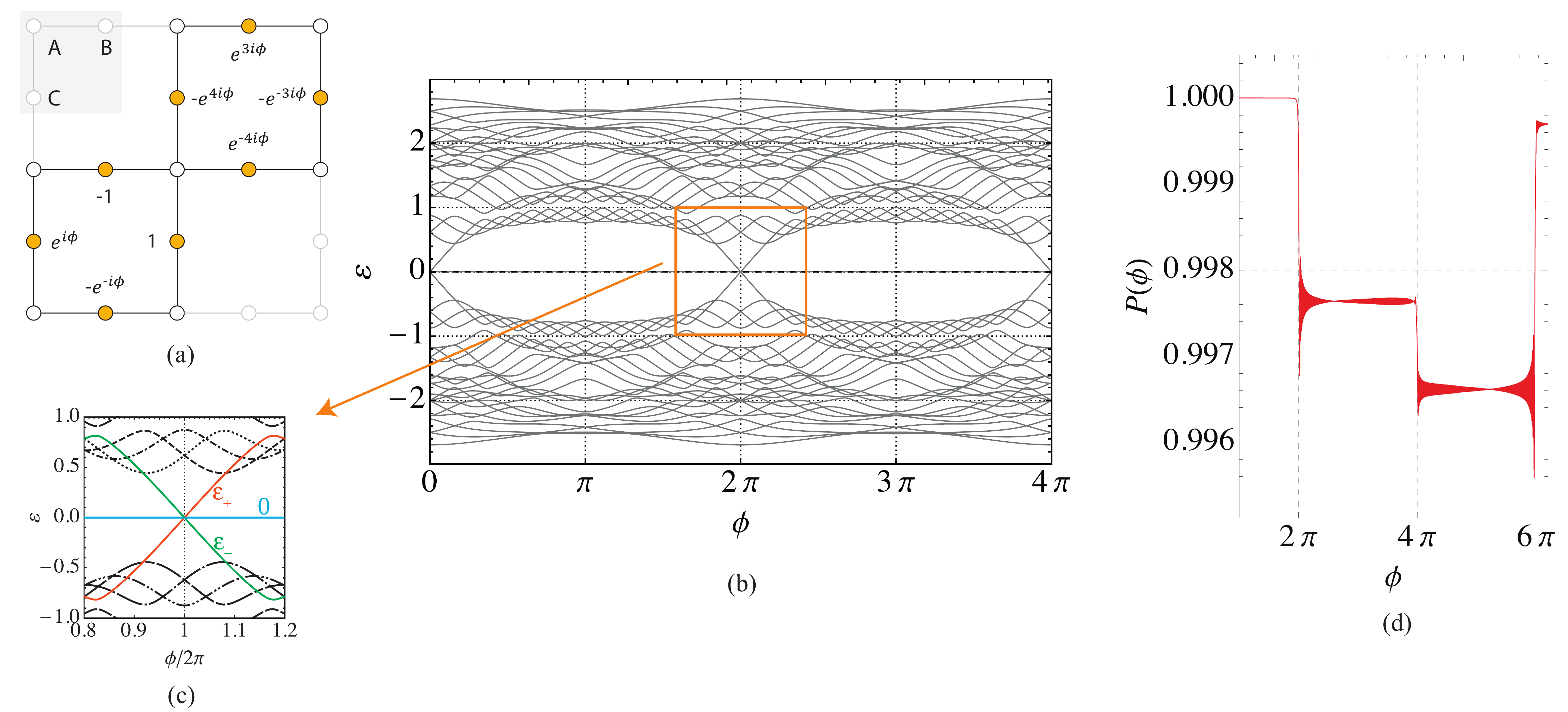}
\caption{(a) Localized state in two contiguous plaquettes of the Lieb lattice with one common site, for the symmetric gauge $\mathbf{A} = \frac{B}{2} (-y,x,0)$. Labeled circles represent finite wavefunction amplitudes, and the remaining sites are nodes of the wavefunction. (b) Energy spectrum of the Hamiltonian $H$ (see text) as a function of the magnetic flux through a plaquette, $\phi$, for the finite Lieb lattice with 4$\times$4 plaquettes. (c) Closeup of an intersection point between the flat band and two dispersive states, which we call the two $\varepsilon$ states. (d) Square of the absolute value of the projection of an evolving state $\vert \psi (t) \rangle$ onto the localized subspace of the eigenvectors of $H$, as a function of $\phi$. The lattice comprises 4 plaquettes (2 in each direction $x$ or $y$). The initial state is of the form of (a), with $\phi = \pi$. The time evolution is due to the linear change of the magnetic flux, $\phi(t)= \omega t$, $\omega = 2\pi\times10^{-5}$ and $(x_0,y_0) = (-4,-4)$.}
\label{fig-energy_phi}
\end{figure*}

Let us consider the Lieb tight-binding model without magnetic flux. The Lieb lattice can be obtained from a traditional two-dimensional (2D) square lattice by removing one quarter of its atoms in a regular pattern (see Fig.~\ref{fig-energy_phi}a). It is comprised of three sublattices (A, B and C). The eigenvalues of the nearest-neighbor TB model of this lattice (with unitary hopping constant) consist of three energy bands, one of which is flat, with zero energy\cite{Nita2013,Mukherjee2015}. The energy of the dispersive bands has the form $E_{\pm}(\mathbf{k}) = \pm 2 \sqrt{\cos^2 \frac{k_x}{2} + \cos^2 \frac{k_y}{2}}$, where (for periodic boundary conditions) $k_{\alpha} = 2 \pi n_{\alpha}/L_{\alpha}$, with $n_{\alpha} = 1,\cdots,L_{\alpha}$, and $L_{\alpha}$ is the number of unit cells in the $\alpha$ direction. The total number of unit cells is $N \equiv L_x L_y$. The flat band is a high-degeneracy eigenspace composed of localized states (these states remain eigenstates when the system size is increased, implying that the respective probability density distribution is localized in a region of the lattice). 

On an infinite Lieb lattice, the dispersive bands have Dirac cones that touch the flat band at the point $\mathbf{k} = (\pi,\pi)$. On a finite periodic Lieb lattice, this point is only allowed if both $L_x$ and $L_y$ are even. When this is the case, the degeneracy of the zero-energy subspace is $N+2$. The number of localized states (degeneracy of the flat band) is $N+1$. The remaining zero-energy state is the eigenstate of the dispersive bands that is located in the Dirac point.
A localized state is also located at the Dirac point, effectively creating a two-state subspace that is degenerate for both energy and
$\mathbf{k}$.
\footnote{The variable $\mathbf{k}$ should not be designated as momentum in the case of open boundary conditions, since the respective eigenstates are standing waves.}
In contrast, on a Lieb lattice with open boundary conditions, the flat band is $N$-fold degenerate in the absence of magnetic flux.
Note that localized states span only over B- and C-type atoms of the lattice, but the zero-energy dispersive state (corresponding to the Dirac point) spans uniquely A-type atoms. This state has finite amplitude at A-type atoms at the edges of the lattice, and consequently it is no longer an eigenstate if more plaquettes are added. The lower and upper dispersive bands involve all three sublattices A, B, and C. An important characteristic of the Lieb TB model is that in the presence of magnetic flux the flat band remains flat, albeit with degeneracy $N-1$, that is, even in the presence of magnetic flux, one has localized eigenstates of the TB Hamiltonian induced by the wavefunction destructive interference associated with the particular Lieb geometry.
\footnote{The decrease in the degeneracy can be justified by the fact that, without flux, a localized state can occupy only one plaquette, while with flux, at least two plaquettes are required, which necessarily reduces the number of localized states.}

To include a magnetic field in the model, we must consider the Peierls phase gained by the electron when it hops between lattice sites, $\theta_{ij} = \frac{\pi}{\phi_0} \int_{i}^{j} \mathbf{A} \cdot d \mathbf{l}$, where $i$ and $j$ label the $(x,y)$ coordinates of the initial and final sites, respectively, $\mathbf{A}$ is the vector potential, and $\phi_0 = h/(2e)$ is the magnetic flux quantum. Assuming the symmetric gauge, $\mathbf{A} = \frac{B}{2} (-(y-y_0),x-x_0,0)$, where $B$ is the magnitude of the magnetic field and $(x_0,y_0)$ is the center of the vector potential relative to the center of the lattice $(0,0)$, the Lieb TB Hamiltonian in the presence of magnetic flux is obtained by applying the Peierls substitution to the standard TB Hamiltonian, and is given by\cite{Harris1989}
\begin{equation}
\begin{split}
&H = - \hspace{-7pt} \sum\limits_{\text{all A sites}} \hspace{-7pt} \left( e^{-i \phi\frac{ (x-x_0)}{8}} B^\dag_{x,y+1}
 + e^{i \phi\frac{ (x-x_0)}{8}} B^\dag_{x,y-1} \right. \\
 +& \left. e^{i \phi\frac{ (y-y_0)}{8}} C^\dag_{x+1,y} + e^{-i \phi\frac{ (y-y_0)}{8}} C^\dag_{x-1,y} \right) A_{x,y} + \text{H.c.},
\label{eq-H_phi}
\end{split}
\end{equation}
where $\phi = 4B\pi/\phi_0$ is the normalized magnetic flux. Open boundaries are introduced considering only the set of the previous hopping terms within the boundaries of our lattice. 
The eigenvalues of the Lieb TB Hamiltonian as a function of $\phi$ are shown in Fig.~\ref{fig-energy_phi}b, which includes a zoomed-in energy-crossing point, Fig.~\ref{fig-energy_phi}c (see also Refs. \onlinecite{Aoki1996,Nita2013, Goldman2011}).
A double Hofstadter butterfly arises in intervals of $2 \pi$. 

The introduction of magnetic flux opens gaps between the bands, and two states, $\vert \varepsilon_+ \rangle$ and $\vert\varepsilon_- \rangle$ (whose energies obey the relation $\varepsilon_+ = - \varepsilon_-$), leave the flat band (see Fig.~\ref{fig-energy_phi}c). These two states arise (up to zeroth order on the flux) from a combination of the two states in the zero-flux Dirac point, one dispersive and one localized. In states $\vert \varepsilon_+ \rangle$ and $\vert\varepsilon_- \rangle$, the electron has equal probability of being at sublattices A or B/C. All A sites have the same probability of occupation, but for the B/C sites the probability increases quasi-exponentially as we move away from the center. This means that the overlap between a localized state and the $\vert \varepsilon_\pm \rangle$ states is stronger the closer the localized state is to the edge of the lattice. In these two states, the phase difference between nearest-neighbor sites is $\pi/2$ as we move clockwise in one of the $\varepsilon$ states and anti-clockwise in the other. This can be interpreted as the two states having opposite angular momenta which, when coupled to the applied magnetic field, confers them symmetric energies at the energy crossing instants.

The zero-energy crossing at zero flux (or more generally, at multiples of the flux quantum) is rather particular. At zero flux and assuming $N_x=N_y$, the Lieb lattice shares the $C_{4v}$ symmetry of the square lattice and therefore, one expects a zero-flux energy spectrum with non-degenerate (double degenerate) states which are even (odd) under the $C_2$ rotation. However, the zero-energy crossing involves the $\vert \varepsilon_\pm \rangle$ states  which, when $(x_0,y_0)=(0,0)$, are even under the $C_2$ symmetry. The introduction of flux lowers the symmetry of the lattice (or better, of the respective tight-binding Hamiltonian), from $C_{4v}$ to $C_{4}$, if we consider a vector potential which has $C_4$ symmetry (as in the case $(x_0,y_0)=(0,0)$), or to a lower symmetry otherwise.

Let us now consider a time-dependent magnetic flux. The time evolution of a localized eigenstate $\vert \psi (0) \rangle$ of the Lieb lattice is given by the time-dependent Schroedinger equation, $i \frac{\text{d}}{\text{d} t} \vert \psi \rangle = H \vert \psi \rangle$, so that $\vert \psi (t+\text{d}t) \rangle = e^{-iH(t) \text{d}t} \vert \psi (t) \rangle$. We considered, as initial state, a localized eigenstate of the Hamiltonian of the Lieb lattice at a certain magnetic flux, and numerically studied its evolution due to a time-dependent Hamiltonian $H (t)$ representing the slow linear change of the magnetic flux, $\phi(t)=\omega t$, where $\omega$ is the angular frequency of the Peierls phase.
One should again note that a slowly-changing time-dependent vector potential implies a very small electric field\footnote{In the context of atomic physics, a gauge such that the electric field results from a time-dependent vector potential is designated a velocity gauge.}.

Starting with the localized state of the Lieb lattice in Fig.~\ref{fig-energy_phi}a, placed at the center of a Lieb lattice with $2 \times 2$ plaquettes, with $\phi(t_0) = \pi$, the projection of $\vert \psi (t) \rangle$ onto the localized subspace of $H (t)$ is shown in Fig.~\ref{fig-energy_phi}d. This projection is given by $P(t) = \sum_i \langle 0_i \vert \psi (t) \rangle^2$, where the summation is over all zero-energy eigenstates, $\vert 0_i \rangle$, of $H$. The fast oscillations with modulated amplitude and a larger-scale staircase behavior, seen in Fig.~\ref{fig-energy_phi}d, are also observed in larger lattices. 
Note that the sum of the state projections onto the localized basis is a particular state $\vert\tilde{0}\rangle $ of the subspace of localized states, $\vert\tilde{0}\rangle = \sum_i \langle 0_i \vert \psi (t) \rangle \vert 0_i\rangle $.

The fact that the localized component exhibits a staircase behavior is a consequence of $\vert \psi (t) \rangle$ acquiring or losing dispersive component in the two $\varepsilon$ states (see Fig.~\ref{fig-energy_phi}c) whenever the $\varepsilon_\pm$ energies cross the flat band (which occurs periodically, at $\phi = 2\pi n$), in accordance with the adiabatic theorem.
This pattern can be successfully reproduced using a three-level toy model, as we show in the next section.
We can justify this pattern by analyzing the equation of evolution of $\vert \psi (t) \rangle$ in the time-dependent eigenbasis $\{\vert \varepsilon_i (t) \rangle \}$ of the Hamiltonian, where we can write $\vert \psi (t) \rangle= \sum_i \alpha_i(t) \vert \varepsilon_i (t) \rangle $. This leads to the equation
\begin{equation}
\text{d}\Psi/\text{d}t = (-iH_d + D)\Psi, 
\label{eq-Evo_equation}
\end{equation}
where $\Psi(t)=\{\alpha_i(t)\}$ is the column vector of the components of $\vert \psi (t) \rangle$ in the eigenbasis $\{\vert \varepsilon_i(t) \rangle \}$. In the equation above, $H_d$ is the diagonalized Hamiltonian matrix and $D_{ij} \equiv \frac{\text{d}\langle \varepsilon_i \vert}{\text{d}t} \vert \varepsilon_j \rangle = \frac{\text{d}\phi}{\text{dt}} \frac{\text{d}\langle \varepsilon_i \vert}{\text{d}\phi} \vert \varepsilon_j \rangle $ where $\phi$ is the magnetic flux that will be varied quasi-adiabatically over time. This implies $\text{d}\phi/\text{dt}$ is small enough that the energy differences between states $i$ and $j$, $\Delta \varepsilon_{ij}$, are (mostly) much greater than the elements that couple those states, $D_{ij}$, so that the matrix $D$ can be considered a perturbation of the system. In this case, the evolution is mostly determined by the diagonalized Hamiltonian, resulting in constant $\vert \alpha_i \vert$ of the evolving state. However, if $\Delta \varepsilon_{ij}$ is zero at some instant $t_1$, the element $D_{ij}$ will dominate the evolution on a finite interval around $t_1$ for any (finite) choice of $\text{d}\phi/\text{dt}$, resulting in a permanent exchange of component weight between states $i$ and $j$.
\section{Three-level toy model}

\begin{figure*}[htbp]
\includegraphics[width=.8 \textwidth]{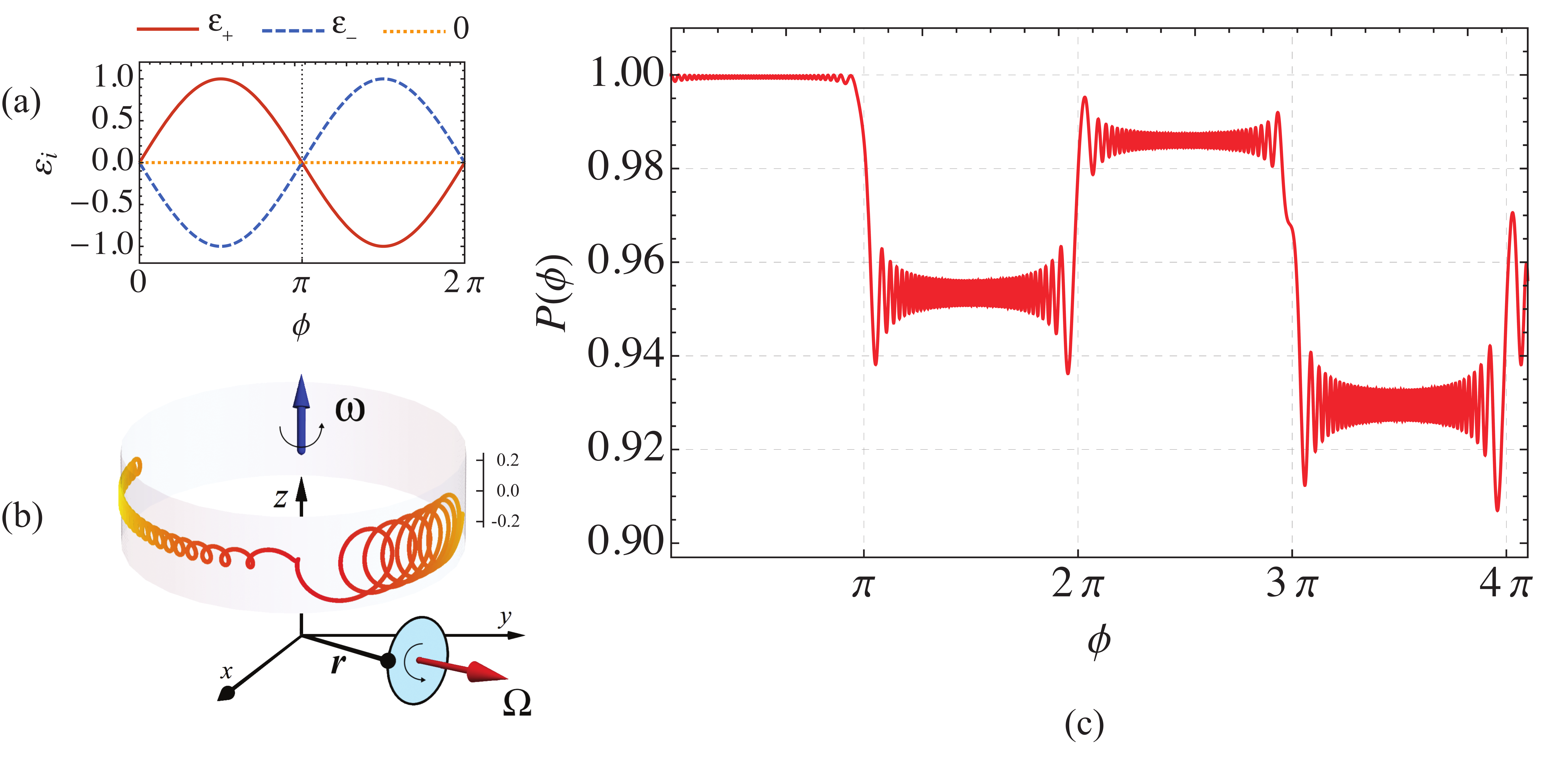}
\caption{(a) Energy spectrum of the three-level toy Hamiltonian $H_3$ as a function of $\phi = \omega_1 t$. The spectrum includes three energy bands, with energies $0$ and $\varepsilon_\pm = \pm \sin(\omega_1 t)$. (b) Illustration of the solution $\vert \psi (t) \rangle$ of the Schroedinger equation describing our three-level toy model as a classical mechanics precessing position vector $\mathbf{r} (t) = \vert \psi (t) \rangle$. The vectors here have the same meaning as in Eq.~\ref{eq-precession}. The trajectory described by the vector $\mathbf{r} (t)$ is the curly orange line and results from solving Eq.~\ref{eq-precession}, with $\omega_1 = \omega_2 = 2\pi/1000$, and initial condition $\mathbf{r} (\omega_1 t=\pi/10) = (\cos(\pi/10),\sin(\pi/10),0)$, i.e., a purely localized state. Note that the direction of precession changes whenever the sign of $\varepsilon(t)$ changes. (c) Projection of the evolving state $\vert \psi (t) \rangle$ onto the zero-energy state of the toy Hamiltonian $H_3$ as a function of $\phi$. The parameters considered are the same as in (b).}
	\label{fig-E_3x3}
\end{figure*}

In this section, we show that the fast oscillations with modulated amplitude and the larger-scale staircase behavior of the localized component described in the previous section can be understood considering a simple three-level system. 
The three-level toy Hamiltonian, before basis rotation, comprises one zero-energy eigenstate and two finite-energy ones,
\footnote{One could also consider a toy system of only two levels corresponding to one zero-energy state and one finite energy state. However, such a toy model would fail to fix the near-zero energy of the evolving state which is observed in flat-band systems. The evolving state needs to be allowed to mix with two other states with symmetric energies, lest the energy change.}
and its matrix representation at time $t$ can be
\begin{equation}
\tilde{H}_3 = \left(
\begin{array}{ccc}
0 & 0 & 0 \\
0 & 0 & \varepsilon(t) \\
0 & \varepsilon(t) & 0
\end{array}
\right) ,
\end{equation}
with eigenvalues $0$ and $\varepsilon_\pm = \pm \varepsilon(t)$, where $\varepsilon(t) = \varepsilon_0 \sin(\omega_1 t)$. The eigenvalues are therefore distinct for all times except $\omega_1 t = n \pi$, with $n \in \mathbb{Z}$ (see Fig.~\ref{fig-E_3x3}a), an effective simplification of the zero-energy crossing instants which occur in the case of the Lieb lattice (see Fig.~\ref{fig-energy_phi}c).

Additionally, a rotation of the eigenbasis of the toy Hamiltonian should be considered. We choose a simple case of a unitary transformation, the rotation matrix $U$ about the $z$ axis, with angular velocity $\omega_2$, so that the zero-energy state mixes with the other two states. The full toy Hamiltonian is $H_3 = U \tilde{H}_3 U^\dag$.
The zero-energy state of $H_3$ at time $t$ is $\vert 0(t) \rangle = (\cos(\omega_2 t),\sin(\omega_2 t),0)$, and the (not normalized) eigenstates with energies $\varepsilon_{\pm}$ are $(\sin(\omega_2 t),\cos(\omega_2 t),\pm 1)$. Other rotations could be considered by substituting $\omega_2 t$ with the appropriate time-dependent functions. We comment the case of a general unitary operator $U$ (with complex matrix entries) in the next section.

The evolution of a quantum state due to this Hamiltonian can then be studied by solving the time-dependent Schroedinger equation, $H_3 \vert \psi (t) \rangle = i \frac{\text{d}}{\text{d}t} \vert \psi (t) \rangle$. One can then plot the projection of the solution vector, $\vert \psi (t) \rangle = (x(t),y(t),z(t))$, onto the zero-energy state, $P_0(t) = \vert \langle \psi (t) \vert 0(t) \rangle \vert^2$. A numerically-obtained plot of $P_0(t)$ is shown in Fig.~\ref{fig-E_3x3}d, where we assumed $\omega_1 = \omega_2$ for simplicity. The energy of the evolving state, $\langle \psi (\varepsilon) \vert H_3 \vert \psi(\varepsilon) \rangle$, was found to remain zero (lower than $10^{-16}$ in our numerical calculations) at all times during the evolution.

The solution of this system can actually be visualized in 3D, as a position vector. First, since $x(t)$ and $y(t)$ have zero imaginary part, and $z(t)$ has zero real part, we make all three components purely real, by replacing $z$ with $iz$ and consider the position vector, $\mathbf{r}(t) = (x(t),y(t),z(t))$. That turns the Hamiltonian into a skew-symmetric matrix, which can be identified as the matrix multiplication form of a vector cross product. Finally, the Schroedinger equation $H_3 \vert \psi (t) \rangle = i \hbar \frac{\text{d}}{\text{d}t} \vert \psi (t) \rangle$ becomes a classical mechanics precession equation,
\begin{equation}
\mathbf{\dot{r}}(t) = \mathbf{\Omega}(t) \times \mathbf{r} (t),
\label{eq-precession}
\end{equation}
with $\mathbf{\Omega} (t) = \frac{\varepsilon_0 \sin(\omega_1 t)}{\hbar} \left[ \cos(\omega_2 t),\sin(\omega_2 t),0 \right]$. This is the equation of motion that describes the precession of a vector $\mathbf{r}(t)$ around the vector $\mathbf{\Omega}(t)$, which in turn rotates about the $z$ axis (Fig.~\ref{fig-E_3x3}b). Thus, in this classical perspective, $\omega_1$ is responsible for the change of the length of $\mathbf{\Omega} (t)$ over time and $\omega_2$ gives the rotation of $\mathbf{\Omega} (t)$ around the $z$ axis.

Because the velocity, $\mathbf{\dot{r}}(t)$, is orthogonal to $\mathbf{r}(t)$ at all times, the norm of $\mathbf{r}(t)$ is kept constant. The vector $\mathbf{\Omega}(t)$ is the zero-energy eigenstate of the Hamiltonian, multiplied by $\varepsilon_0 \sin(\omega_1 t)/\hbar$. The zero-energy component of $\mathbf{r}(t)$ is therefore proportional to the cosine of the angle between $\mathbf{r}(t)$ and $\mathbf{\Omega}(t)$. At the energy intersection points ($\omega_1 t = n \pi$), the velocity goes to zero and $\mathbf{r}(t)$ moves more slowly. However, at these points, $\mathbf{\Omega}(t)$ continues to rotate at the same angular velocity so that, naturally, the angle between the two vectors changes considerably, inducing the sudden increases or decreases in the localized component that can be seen in Fig.~\ref{fig-E_3x3}d. Away from the intersection points, $P_0(t)$ displays rapid oscillations which reflect the precession of $ \mathbf{r}$ around $\mathbf{\Omega}$, as well as the rotation of $\mathbf{\Omega}$. 

In the rotating frame of reference that follows the zero-energy eigenstate, the precession vector acquires a component in the $z$ direction, $\mathbf{\Omega} = [\varepsilon_0 \sin(\omega_1 t)/\hbar,0,-\omega_2]$. The instantaneous frequency of the rapid oscillations of $P_0(t)$ observed in Fig.~\ref{fig-E_3x3}d equals the norm of the precession vector in the rotating frame, $\Omega = \sqrt{(\varepsilon_0 \sin(\omega_1 t)/\hbar)^2+\omega_2^2}$ (however, if $\omega_2$ is zero, no oscillations will be observed). Indeed, a WKB-like approach can be used to find an approximate solution of the three-level model in the regime $ \lvert \varepsilon(t)/\hbar \rvert \gg \lvert \omega_2 \rvert \Rightarrow \Omega(t) \approx \varepsilon(t)/\hbar$. The approximate solution for the localized projection $\langle \psi(t) \vert 0(t) \rangle$ between two consecutive zero-energy crossing instants is 
\begin{equation}
c \frac{\varepsilon_0}{\Omega}\sin(\omega_1 t) + \sqrt{1-c^2} \frac{\omega_2}{\Omega} \cos\left(-\frac{\varepsilon_0}{\omega_1}\cos(\omega_1 t) + \theta_0 \right),
\label{eq-Osci_aprox}
\end{equation}
where $-1 < c < 1$ and $\theta_0$ are initial condition parameters. This is still a valid equation even if we used a more general rotation of the eigenstates, i.e. if the $z$ component of the rotating frame precession vector, $-\omega_2$, was substituted by any time-dependent function. This means that the amplitude of the high frequency oscillation of $\lvert \langle \psi(t) \vert 0(t) \rangle \rvert^2$ will be approximately given by $\lvert \omega_2/\Omega \rvert^2$.

\section{Electric field symmetry}

In the previous section, we showed that a simple three-level system can reproduce the basic features of an evolving localized state in the Lieb lattice, namely the fast oscillations with modulated amplitude of the localized component and the larger-scale staircase behavior.
The precession of $\mathbf{r}(t)$ in the case of the toy model implies that the oscillations observed in the case of the Lieb lattice can be qualitatively interpreted as a precession of the evolving state around a state which is approximately the state $ \vert\tilde{0}\rangle$ defined in section \ref{sec:lieb}, but also has small $\vert \varepsilon_\pm \rangle$ components (and even smaller components on other itinerant states).
Furthermore, the nodes and antinodes in the amplitude of the oscillations at each step of the staircase (this effect is more explicit for larger lattices) observed in Fig.~\ref{fig-energy_phi}d can also be reproduced by the toy model by tweaking $\omega_2$, as it modulates the amplitude of the oscillations. This implies that, if in a certain instant $\omega_2$ is zero, then a node will be observed in the amplitude of the oscillations (see Eq.~\ref{eq-Osci_aprox}).


In the case of the Lieb lattice, and in analogy with the three-level system, a rotation between the $\vert \varepsilon_\pm \rangle$ states and the localized states occurs as time evolves. Since this rotation occurs as the magnetic flux is increased, one expects the rotation to be proportional to the time derivative of the Hamiltonian. Since the time dependence of the Hamiltonian is present only in the vector potential, the rotation reflects the existence of an electric field. But does any electric field generate such a rotation? Or equivalently, is it possible to define a time-dependent vector potential such that the corresponding electric field does not cause the step-like behavior of the localized component of the evolving state? The answer lies in the relative symmetry of the lattice and the vector potential. As mentioned above, a slowly-changing time-dependent vector potential implies a very small electric field, given by $\mathbf{E} =-\partial \mathbf{A}/\partial t$. In this work, we used the symmetric gauge, $\mathbf{A} = \frac{B(t)}{2} (y_0-y,x-x_0,0)$ with a linear time dependence of the magnetic field, $B(t) = \omega t \phi_0/(4\pi)$. The specific case where $(x_0,y_0)=(0,0)$, meaning the center of the gauge is the same as that of the lattice, leads to the same magnetic field, but a different electric field, and the step-like behavior vanishes. In this case, both the system and the electric field possess rotation invariance at the center of the lattice and therefore, eigenstates of $H$ have odd or even parity in relation to the center of the lattice. In particular, one can choose a Hamiltonian eigenbasis for the localized states, $\{\vert 0_i(t) \rangle\}$, such that all states have a defined parity.


In a time-dependent evolution, the transition rate between eigenstates $\vert \varepsilon_+(t) \rangle$ and $\vert 0_i(t) \rangle$ is given by $D_{+0_i}(t) = \langle \frac{\text{d}\varepsilon_+}{\text{d}t} \vert 0_i(t) \rangle$, if $\varepsilon_+ \neq 0$. Using the $C_4$ symmetries of both states, one can show that the transition rate at the crossing points $\phi = 2\pi n$ is proportional to the amplitude of the uniform component of the electric field, $D_{+0_i}(\phi = 2\pi n) \propto \omega\sqrt{x_0^2+y_0^2}$ (see Ref.~\onlinecite{Maceira2016} for more details).
As stated by the adiabatic theorem, a slow time evolution may only leave the adiabatic regime if an energy difference of zero is met. However, as discussed in the analysis of Eq. \ref{eq-Evo_equation}, the adiabatic regime is abandoned when the matrix elements of $D$ are sufficiently larger than the energy differences between the respective states. Since $D_{+0_i} = 0$ exactly at the same time as the energy difference is zero, there is not a finite time interval around the crossing instants where $D_{+0_i} \gg \Delta E_{+0_i}$, so that adiabaticity is not lost even though the energy levels meet, causing the absence of the staircase behavior when the electric field shares its center with the lattice. 

\section{Conclusion}

In conclusion, we have studied the slow time evolution of localized states of the Lieb lattice with increasing magnetic flux. A curious step pattern of the localized component has been found and we have shown that this behavior can be interpreted as a precession movement of the evolving state around a time-dependent vector with a large localized component and a much smaller dispersive component.
The small dispersive component changes sharply at the energy-crossing points and corresponds mainly to two eigenstates of the Hamiltonian whose energy periodically crosses the zero-energy line. 
We have shown that this behavior can be understood considering a simple three-level toy model consisting of a Hamiltonian with three time-dependent eigenstates, such that one of them has constant zero energy and the other two periodically cross the zero-energy line. 

This behavior should also occur due to perturbations that, similarly to the magnetic and electric fields, lift the C$_{4v}$ symmetry of the Lieb lattice. We also expect that other flat band systems display the same features. For example, similar behavior is found in the $AB_2$ chain\cite{Lopes2011}, which is bipartite and also has a flat band which is robust against the application of a magnetic field. 

Concerning the experimental observation of the physics described in this paper, the step pattern may be observed in Lieb optical lattices under time-dependent perturbations or in Lieb photonic lattices, using a spatial modulation of the properties of the waveguide\cite{Longhi2006} to replicate the time-dependent magnetic field\cite{Golshani2014,Flach2016}, by measuring the light intensity at an A-type waveguide over its length, we can approximately measure the itinerant component over time. 

\begin{acknowledgments}
\appendix
J. D. Gouveia acknowledges the financial support from the Portuguese
Institute for Nanostructures, Nanomodelling and Nanofabrication (i3N) through the grant BI-14/I3N-LA25/MAIO15 (1307/1309-2015).
R. G. Dias acknowledges the financial support from the Portuguese
Science and Technology Foundation (FCT) through the program PEst-C/CTM/LA0025/2013. R. G. D. thanks the support by the Beijing CSRC.
\end{acknowledgments}

\bibliographystyle{unsrt}
\bibliography{biblio}

\end{document}